# HybridNetSeg: A Compact Hybrid Network for Retinal Vessel Segmentation

**Ling Luo,**[1*] **Dingyu Xue,**[1] **Xinglong Feng,**[1]

[1] Northeastern University, College of Information Science and Engineering, Shenyang, China, 110819

**Abstract**. A large number of retinal vessel analysis methods based on image segmentation have emerged in recent years. However, existing methods depend on cumbersome backbones, such as VGG16 and ResNet-50, benefiting from their powerful feature extraction capabilities but suffering from high computational costs. In this paper, we propose a novel neural network (HybridNetSeg) dedicated to solving this drawback while further improving overall performance. Considering deformable convolution can extract complex and variable structural information, and larger kernel in mixed depthwise convolution makes contribution to higher accuracy. We have integrated these two modules and propose a Hybrid Convolution Block (HCB) using the idea of heuristic learning. Inspired by the U-Net, we use HCB to replace a part of the common convolution of the U-Net encoder, drastically reducing the parameter count to 0.71M while accelerating the inference process. Not only that, we also propose a multi-scale mixed loss mechanism. Extensive experiments on three major benchmark datasets demonstrate the effectiveness of our proposed method. Code is available at: https://github.com/JACKYLUO1991/HybridNetSeg

*Ling Luo, E-mail: lingluo@stumail.neu.edu.cn

## 1 Introduction

Pixel-level semantic segmentation, as one of the three major directions of computer vision has made remarkable progress in recent years. Nowadays, this technology has been successfully applied to medical imaging diagnosis, e.g., cell detection[1,2,3], bood vessel segmentation[4,5] and optic disc segmentation[6,7]. In terms of retinal vessel segmentaion task, efficient and accurate segmentation results is conductive to diagnosing ophthalmic diseases which may lead to diabates, hyoertension and other diseases. Due to the emergence of deep convolutional neural networks (DCNNs), researchers abandon the traditional hand-crafted features and switch to DCNNs to extract features automatically, which can improve the performance of the model to some extent. Among the existing massive segmentation models, U-Net[8] is the most representative of this field which consists of a U-shaped encoder-decoder. In particular, this model is capable of extracting contextual information in contrast to patch-based models. Most of the later methods are derivative versions of original U-Net. However, usually these methods do not achieve a balance of accuracy, time complexity and memory footprint.

In the process of model deployment, not only the segmentation quality, but also model size and model forward pass calculated amount are important factors. On the other hand, considering the irregularity shape of the retinal blood vessels, the generalization of the model is usually poor. A common way to tackle it is to apply data augmentation, which does not solve this problem essentially from the model itself. Herein, we propose a novel compact model that reduces the FLOPs of the model while maintaining accuracy.

In this paper, we integrate the design advantages of existing neural network modules and develop a novel model called HybridNetSeg that embeds mixed depthwise convolution[22] (MixConv) and deformable convolution[21] (DCN). In our HybridNetSeg, MixConv combines different sizes of convolution kernels in a single operator in order to capture various patterns. DCN has a better learning ability for deformable targets than common convolution while capturing more structual information. Additionally, a mixed loss function is used to smooth the loss and further improve the performance of the model.

The **main contributions** of this paper can be summarized as follows:

- We propose a Hybrid Convolution Block, which combines DCN and MixConv to improve the performance of the model while capturing edge details.
- We present a novel loss function named Mixed loss that makes the gradient smoother during the back propagation.
- We give the most comprehensive evaluation metrics on publicly benchmark datasets.

- We show that our HybridNetSeg achieves state-of-the-art performance on three retinal vascular datasets. Further evaluation of the inference time confirms that the solution can be applied to edge and embedded devices.

## 2  Related Work

**Semantic segmentation:** DCNNs have gradually become the most popular paradigm in the field of semantic segmentation, eliminating the cumbersomeness of traditional handcrafted features. As a sub-direction of the semantic segmentation, the retinal vessel segmentation task has been widely developed in recent years thanks to high-quality network architecture[9,10,11]. The milestone work of U-Net[8] explores a network that can be trained end-to-end in the case of very few images, consisting of a contracted path to capture context and symmetric expanded path, which is mainly applied to pixelwise classification. Furthermore, the symmetrical network specific layers are concatenated with *shortcut*[12] to fuse features of different scales.

Most of the current medical image segmentation models are variants of U-Net. Fu *et al*.[13] takes advantage of conditional random fields (CRFs) to model the long-term dependencies between pixels. M-Net[7] ulilizes image pyramid mechanism to achieve multiple level receptive field sizes. Tim *et al*.[14] adopts light-weight network structure MobileNet v2 as backbone, introduces contracting bottleneck blocks in the decoder part at the same time, making the model suitable for embedded devices. Nabil *et al*.[15] analyzes the possible segmenatic gaps between the corresponding levels of the U-Net encoder-decoder, and proposes a '*Res Path*' module. Gu *et al*.[5] presents dense atrous convolution block and residual multi-kernel pooling components in order to capture more high-level features and preserve more spatial information. Yu *et al*.[27] utilizes 'Edge Guidance' module that embeds edge detection into the network to extract sufficient edge information. Recently, Zhang *et al*.[16] exploits visual attention mechanism and edge guided filter to ameliorate boundary detail information.

**Loss function:** The most common objective function of image segmentation is pixelwise cross-entropy. However, it is often confronted with the problem that the performance of the model will drop dramatically when the categories are unbalanced. In terms of binary segmentation tasks, Fausto *et al.*[17] introduces a novel loss function called '*Dice loss*' which is based on *Dice coefficient* in order to relieve extreme imbalance between foreground and background pixels. Similarly to address this problem, *Focal loss*[18] consists of weighted cross-entropy and hard examples of redistribution weights. *Tversky loss*[19] is a generalization of Dice's coefficient, adding a weight to false positives and false negatives.

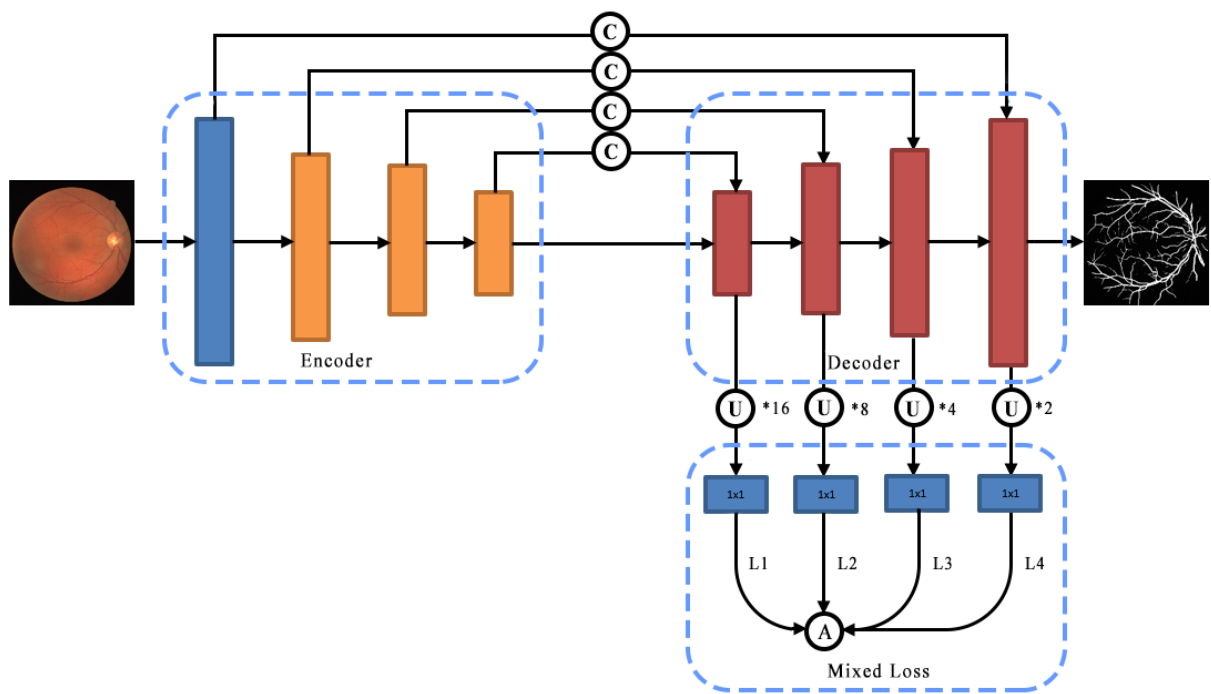

**Fig. 1** The illustration of HybridNetSeg structure. From left to right, the block diagrams of different colors represent common convolution, our proposed hybrid convolution block and contract inverted bottleneck block. Among them, the operator 'C' is adopted for hierarchical fusion between feature maps of the same resolution in the encoder and decoder. 'U' represents a bilinear upsampling operation on the feature map of the corresponding branch. 'A' represents the mixed loss of different scale feature maps.

## 3    Method

Fig. 1 illustrates the overall framework of the proposed HybridNetSeg. The model based on an encoder and decoder structure similar to U-Net, along with hybrid convolution blocks consisting of MixConv and DCN modules. The encoder part consists of one common convolution and three hybrid convolution blocks, each reduces the resolution of the feature map by half. The decoder follows the literature[14], using contractive bottleneck blocks with bilinear unsampling layers to gradually restore the dimensions of the feature map. 'Concatenation' is used between layers of the same dimension to enhance the representation ability of the model. Other than that, a mixed loss function is proposed by weighted averaging and combining between feature maps of various scales. Here, a 1×1 convolution is used to maintain channel consistency. Our proposed HybridNetSeg is detailed as follows.

*3.1 Hybrid Convolution Block*

Hybrid convolution block is a combination of DCNs and MixConvs. For the first time, DCN v1[20] introduced the ability to learn spatial geometry deformation in convolutional neural networks. Unlike traditional regular grid sampling locations, DCN v1 obtains free form deformation of the sampling grids through a parameter-learnable 2D offset. On the other hand, the effective receptive field of the DCN v1 can cover the segmented objects in order to learn more distinguishing features. The formula for DCN v1 under 2D spatial domain is as follows:

$$y(p) = \sum_{i=1}^{N} w_i \cdot x(p + p_i + \Delta p_i) \tag{1}$$

Let $N$ denotes sampling locations of a given standard convolutional kernel, $\omega_i$ and $p_i$ represent the weight and the preset offset (e.g. $p(i) \in \{(-1,-1),(1,1),...,(0,1)\}$) for $i$-th location, repectively. For each location $p$, $x(p)$ and $y(p)$ denote features on the input and output feature maps, respectively. $\Delta p_i$ is the learnable offset obtained by applying a separate convolutional layer on the same input feature map $x$, which has the same spatial resolution and dilation as the current convolutional layer. In addition, since $p + p_i + \Delta p_i$ is fractional, the bilinear interpolation algorithm is applied for calculation. The channel dimension is $2N$, which represents the offsets that can be learned in the x and y directions.

Although DCN v1 can learn a specific object structure compared to a conventional convolution, literature[21] indicates that once this support exceeds the region of interest, the feature may be affected by irrelevant image content. To address this drawback, a modulation mechanism is introduced on the basis of the DCN v1, which achieve more accurate feature extraction by adjusting feature amplitudes from different spatial locations. Here we simply call it DCN v2. Unlike DCN v1, DCN v2 adds a modulation operator to the formula, as follows:

$$y(p) = \sum_{i=1}^{N} w_i \cdot x(p + p_i + \Delta p_i) \cdot \Delta m_i \qquad (2)$$

where $\Delta m_i$ is the modulation factor for $i$-th location, which is limited to 0-1 during the calculation. Taking advantage of these advantages, the DCN v2 module enhances the ability to model geometric deformations in retinal vessels, effectively capturing structural details.

Depthwise convolution can signficantly reduce parameters and computations with a little loss of accuracy, making it the most popular lightweight convolutions. Recent research[22] shows that mixing up multiple kernel sizes in the same convolution op could capture various resolution patterns. More specifically, a larger kernel can obtain large receptive field to improve the accuracy of the model, and small ones can reduce the model parameters for efficiency. For the retinal vessel segmentation task, we gain inspiration from heuristic learning to trade off between computing co-mplexity and efficiency by stacking the above modules. At the same time, considering the superiority of the model[23], the depth separable convolution in the bottleneck of MobileNet v3 is replaced by MixConv, named *MNBlock*.

**Table 1** The encoder structure of HybridNetSeg. s denotes stride. se and t denote squeeze and expansion factor, respectively. k needs to be emphasized, denoting different kernel sizes in a single convolution op. In addition, BN (batch normalization) and nonlinear activation function are not required after DCN.

| Name | Input | Operator | #out | s | k | t | se |
|---|---|---|---|---|---|---|---|
| - | $3\times 512^2$ | Conv2D | 16 | 2 | [3] | - | - |
| Hybrid convolution block | $16\times 256^2$ | DCN | 16 | 1 | [3] | - | - |
| | $16\times 256^2$ | MNBlock | 24 | 2 | [3] | 6 | - |
| | $24\times 128^2$ | MNBlock | 24 | 1 | [3] | 3 | - |
| | $24\times 128^2$ | MNBlock | 40 | 1 | [3,5,7] | 6 | .5 |
| | $40\times 128^2$ | MNBlock | 40 | 1 | [3,5] | 6 | .5 |
| | $40\times 128^2$ | DCN | 40 | 1 | [3] | - | - |
| | $40\times 64^2$ | MNBlock | 80 | 2 | [3,5,7] | 6 | .25 |
| | $80\times 64^2$ | MNBlock | 80 | 1 | [3,5] | 6 | .25 |
| | $80\times 64^2$ | MNBlock | 80 | 1 | [3,5] | 6 | .25 |
| | $80\times 64^2$ | MNBlock | 80 | 1 | [3,5] | 6 | .25 |
| | $80\times 64^2$ | DCN | 80 | 1 | [3] | - | - |
| | $80\times 64^2$ | MNBlock | 80 | 2 | [3,5,7,9] | 6 | .5 |
| | $80\times 32^2$ | MNBlock | 120 | 1 | [3,5] | 3 | .5 |
| | $120\times 32^2$ | MNBlock | 120 | 1 | [3,5] | 3 | .5 |

Here, we simply refer to the modules containing DCN v2 (Hereinafter referred to as DCN) and MixConv as hybrid convolution block. Since the decoder follows the structure[14], we only give the full specification of the encoder in Table 1.

*3.2 Mixed Loss Function*

In the decoding stage, the spatial resolution of the feature map is gradually restored by upsampling. However, through repeated experiments, we observe that the early supervisory information has a significant influence on the result of the segmentation, so a Mixed loss function is proposed:

$$L_{mixed} = \left(1+\frac{1}{n}\right)\sum_{i}^{n} L(upsample(f(i), \ gt)) \tag{3}$$

where $n\,(n=4)$ denotes the number of decoder stage, *upsample* denotes the feature map that restores the original input size after bilinear upsampling, $1\times1$ convolution to adjust the channel number, *gt* is ground-truth. The summed average of the loss functions makes the gradient smoother during back propagation. The specific expression of $L$ is defined as:

$$L = -\frac{1}{n}\sum_{i=1}^{n}(y_i\log(\hat{y}_i) + (1-y_i)\log(1-\hat{y}_i))\cdot w + \left(1 - \frac{1}{n}\sum_{i=1}^{n}\frac{y_i\hat{y}_i}{y_i + \hat{y}_i - y_i\hat{y}_i}\right)\cdot(1-w) \quad (4)$$

where the left side of the formula denotes binary cross entropy loss, and on the right is the Dice loss. The paper[17] points out that the latter can alleviate the imbalance of class to some extent. $n$ denotes the number of pixels in a given image. $\hat{y}\in[0,1]$ and $y\in\{0,1\}$ are the probability of pixels being the foreground and ground-truth, respectively. In our experiments, weight $w$ is empirically set to 0.5.

*3.3 Implementation details*

During training, data augmentation is adopted to enhance the generalization of the model, e.g. random color jitters, random flip, random scale and random shift. According to our model architecture, the input resolution needs to be divisible by 16, so we perform the following resizes:

- **DRIVE**：$512\times512$

- **CHASE_DB1**：$960\times960$

- **HRF**：$784\times1168$ (*Note*: Since the GPU memory is not sufficient (11 GB), HRF image is scaled proportionally)

We train our network from scratch using *AdamW* optimizer with a *learning rate* of 0.001 and a *weight decay* of 0.0005 (default for other parameters). The *mini-batch* size is set to 2 for DRIVE,

CHASE_DB1 and HRF to 1. The whole framework is implemented using Pytorch 1.3. *Early stopping* is another form of regularization used to avoid overfitting, in this experiment, we set *patience* to 30 (the maximum epoch is 500). Additionally, the experimental environment is equipped with a 3.60GHz CPU and a Nvidia GTX 1080Ti graphics card.

## 4 Experiments and Results

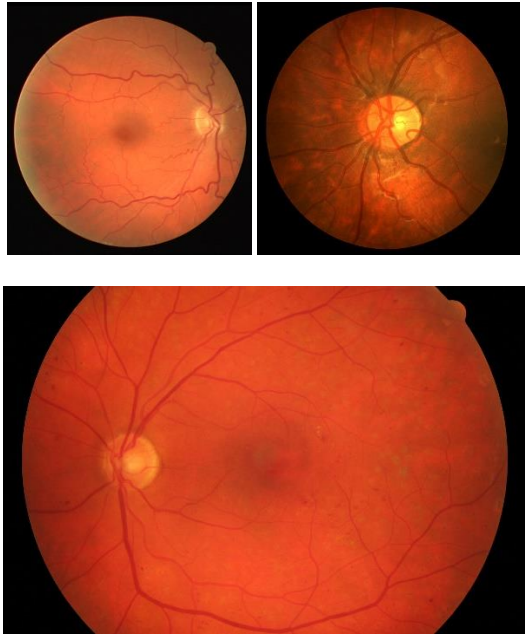

**Fig. 2** Training examples on different datsets. *Top left*: DRIVE (584×565); *Top right*: CHASE_DB1 (960×999); *Bottom*: HRF (2336×3504), the numbers in parentheses represent the corresponding dimensions.

*4.1 Datasets*

To evaluate the performance of our proposed method, we conduct experiments on three benchmark datasets. In the next section, we will briefly introduce these datasets.

**DRIVE.** The DRIVE[24] dataset is manually labeled under the guidance of an experienced ophthalmologist. The dataset contains 40 fine-grained pixelwise annotation photographs in which training images and testing images are equally divided.

**CHASE_DB1.** The CHASE_DB1 dataset is collected by the Child Heart and Health Study in England. We refer to the dataset partitioning scheme in the paper[25], dividing it into 8 images (the first 8 images) for training, and the remaining 20 images for testing.

**HRF.** The HRF dataset contains 45 images, which is divided into 15/15/15 images for healthy patients, patients with diabetic retinopathy and glaucomatous patients. For this dataset, we follow the suggestion of paper[26], where the first five images in each category are used for training and the rest for testing.

An exemplar of each dataset is shown in Fig.2.

*4.2 Evaluation Metrics*

In order to compare the proposed HybridNetSeg with several state-of-the-art algorithms, we give the most comprehensive segmentation evaluation criteria at present. The detailed mathematical expression is as follows:

$$Sen = \frac{TP}{TP + FN} \tag{5}$$

$$Sp = \frac{TN}{TN + FP} \tag{6}$$

$$F1 = \frac{2TP}{2TP + FP + FN} \tag{7}$$

$$Acc = \frac{TP + TN}{TP + FN + FP + TN} \tag{8}$$

where $TP$, $TN$, $FP$ and $FN$ represent the number of true positives, true negatives, false positives, and false negatives, respectively. $F1\text{-}score$ (also known as '$Dice\ score$') is commonly used for performance measurement of medical image segmentation. Although the Intersection over Union ($IoU$) and $Dice\ score$ are monotonically increasing in a mathematical sense, we still give the test results of this metric. In addition, we also employ Area Under Curve ($AUC$) as measurements.

*4.3 Experimental Analysis*

Table 2 Comparison of segmentation results on DRIVE dataset (%)

| Model | Acc | Sen | Sp | Dice | IoU | AUC |
|---|---|---|---|---|---|---|
| U-Net[8] | 0.9681 | 0.7897 | **0.9854** | 0.7941 | 0.6834 | 0.9836 |
| DeepVessel[13] | 0.9523 | 0.7603 | - | - | - | 0.9752 |
| M2U-Net[14] | 0.9630 | - | - | 0.8091 | - | 0.9714 |
| CE-Net[5] | 0.9545 | **0.8309** | - | - | - | 0.9779 |
| ET-Net[27] | 0.9560 | - | - | - | - | - |
| AGNet[16] | 0.9692 | 0.8100 | 0.9848 | - | **0.6965** | 0.9856 |
| **HybridNetSeg** | **0.9695** | 0.8148 | 0.9801 | 0.8041 | 0.6725 | 0.9670 |

Table 3 Comparison of segmentation results on CHASE_DB1 dataset (%)

| Model | Acc | Sen | Sp | Dice | IoU | AUC |
|---|---|---|---|---|---|---|
| U-Net[8] | 0.9723 | 0.7715 | 0.9858 | - | 0.6366 | 0.9837 |
| M2U-Net[14] | 0.9703 | - | - | 0.8006 | - | 0.9666 |
| M-Net[7] | 0.9729 | 0.8089 | 0.9851 | - | 0.6483 | 0.9845 |
| AGNet[16] | 0.9743 | 0.8186 | 0.9848 | - | 0.6669 | 0.9863 |
| **HybridNetSeg** | 0.9732 | 0.8207 | 0.9839 | 0.8004 | 0.6680 | 0.9779 |
| **HybridNetSeg**† | 0.9740 | **0.8217** | 0.9850 | **0.8061** | **0.6758** | 0.9791 |

Table 4 Comparison of segmentation results on HRF dataset (%)

| Model | Acc | Sen | Sp | Dice | IoU | AUC |
|---|---|---|---|---|---|---|
| M2U-Net[14] | 0.9635 | - | - | 0.7814 | - | - |
| **HybridNetSeg** | 0.9672 | 0.7679 | 0.9840 | 0.7826 | 0.6447 | 0.9734 |
| **HybridNetSeg**† | **0.9693** | 0.7624 | **0.9872** | **0.7930** | **0.6581** | **0.9815** |

Table 5 Comparison in terms of model size and inference time. In the case where the input dimension $3 \times 512 \times 512$, results are as follows. **Params**: Number of parameters. **MACs**: Multiply-accumulate operation. **Size**: The space occupied by the model on the disk. **Time:** Inference time on GPU.

| Model | #Params (M) | Flops (GMACs) | Size (MB) | Time (s) |
|---|---|---|---|---|
| U-Net[8] | 31.03 | 192.96 | 119.0 | **0.035** |
| **HybridNetSeg** | **0.71** | **3.52** | **2.8** | 0.037 |

**Comparison on DRIVE.** We compare our algorithm with several state-of-the-art algorithms, as shown in Table 2. From the evaluation results, we can observe that our model has no obvious advantage in low-resolution images, even lower than the classic U-Net. Our hypothesis is that low-resolution images are more sensitive to convolution kernel sizes. Surprisingly, the accuracy of the model is slightly higher than the previous best method, i.e., **96.95**% vs. 96.92%.

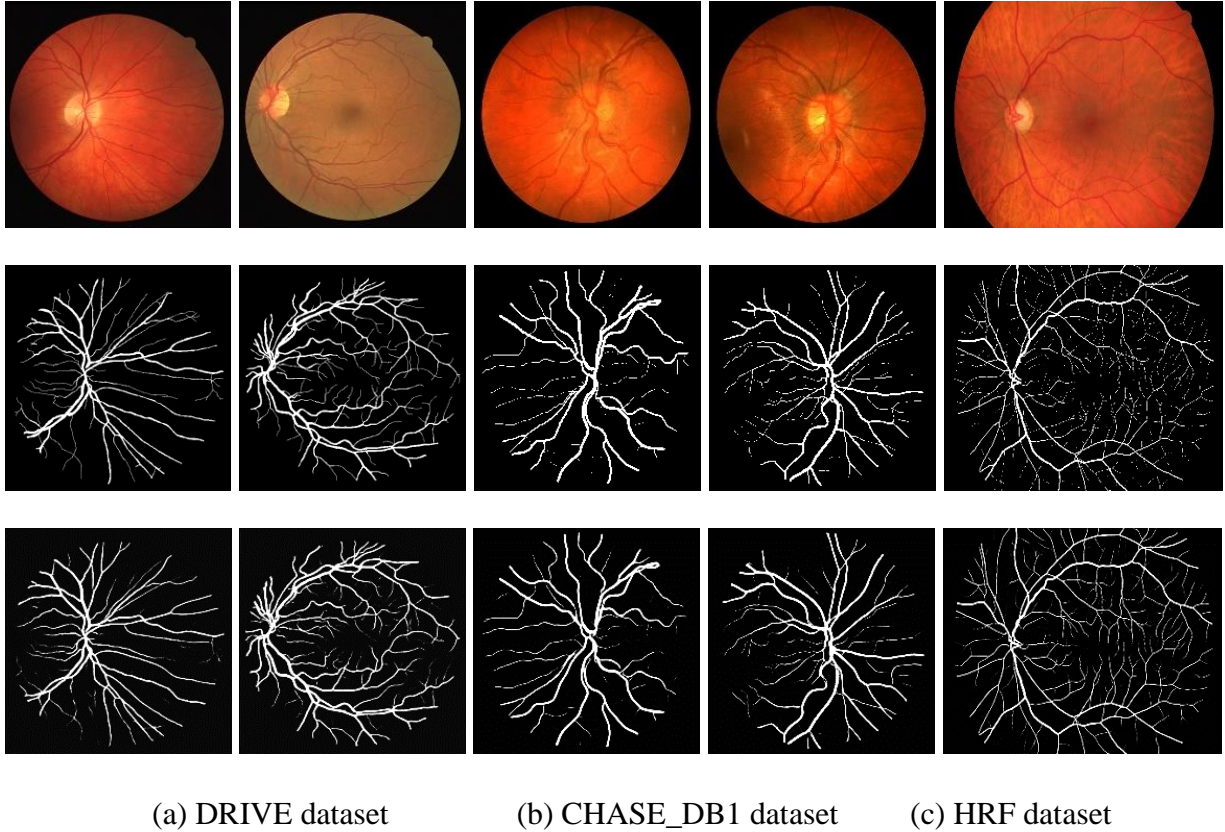

(a) DRIVE dataset      (b) CHASE_DB1 dataset      (c) HRF dataset

**Fig. 3** Qualitative results on benchmark datasets: (a), (b) and (c) represent different datasets, respectively. From top to bottom, RGB image, ground truth and the corresponding probability map are sequentially displayed.

**Comparison on CHASE_DB1 and HRF.** We further evaluate the performance among different methods on high-resolution CHASE_DB1 and HRF datasets. From the results, we can summarize the following viewpoints: Firstly, HybridNetSeg performs better than other methods, which confirms the superiority of Hybrid Convolution Block and Mixed loss. Secondly, unlike DRIVE dataset, HybridNetSeg performs better on CHASE_DB1 and HRF datasets, demonstrating the advantages of MixConv in high-resolution pattern. Lastly, †shows that the pre-trained weights on the DRIVE dataset contribute to the results of the homologous datasets.

**Visualization Results.** The qualitative segmentation results in Fig.3 visually demonstrate the effectiveness of our scheme. From the perspective of edge extraction, we can observe that our model can capture the edge texture details for pleasing results.

**Ablation Study.** In order to justify the effectiveness of HCB and Mixed loss in the proposed HybridNetSeg, we further carry out ablation study on both of CHASE_DB1 and HRF. Since table 3 and table 4 (*HybridNetSeg* vs. *M2U-Net*) already reflect the performance improvement of HCB, and therefore we only analyze the impact of Mixed loss on performance metrics. (yes/no) in Table 6 indicates whether or not mixed loss is used. From the numerical results, we see that almost all metrics have improved with the help of mixed loss. For *Sen* in HRF, the mixed loss one outperforms the other by a large margin (from 0.7367 to **0.7624**), which confirms our hypothes: *Early supervisory information contributed to the accuracy of the model*. Combined with the previous Hybrid Convolution Block, we can indicate the effectiveness of our solution.

Table 6 Ablation study on CHASE_DB1 and HRF

| Metric | CHASE_DB1 | | HRF | |
| --- | --- | --- | --- | --- |
| | yes | no | yes | no |
| Acc | **0.9740** | 0.9732 | **0.9693** | 0.9691 |
| Sen | **0.8217** | 0.8195 | **0.7624** | 0.7367 |
| Sp | **0.9850** | 0.9841 | 0.9872 | 0.9886 |
| Dice | **0.8061** | 0.8007 | **0.7930** | 0.7849 |
| IOU | **0.6758** | 0.6684 | **0.6581** | 0.6481 |
| AUC | 0.9791 | 0.9806 | **0.9815** | 0.9794 |

## 5 Conclusion

In this paper, we present a novel compact model, named HybridNetSeg. HybridNetSeg consists of multiple Hybrid Convolution Blocks which draw the essence of DCN and MixConv. In addition, we introduce multi-scale supervised loss: mixed loss, which significantly improves the performance of the model. Extensive experiments have confirmed that our algorithm can further improve the accuracy while pursuing speed.

*Disclosures*

No conflicts of interest, financial or otherwise, are declared by the authors.

*Acknowledgment*

This research was financially supported by the National Natural Science Foundation of China (No.61773104). The authors state no conflict of interest and have nothing to disclose.
*References*

[1] Xie. W et al., "Microscopy cell counting and detection with fully convolutional regression networks," *Computer methods in biomechanics and biomedical engineering. Imaging & visualization*. **6**(3), 283-292 (2018).

[2] Yao. X et al., "Cell counting by regression using convolutional neural network," in *European Conference on Computer Vision*, *Lecture Notes in Computer Science*, pp. 274-290, Springer, Cham (2016).

[3] Akram. S. U et al., "Cell segmentation proposal network for microscopy image analysis," in *Deep Learning and Data Labeling for Medical Applications*, pp. 21-29, Springer, Cham (2016).

[4] Jin. Q et al,. "DUNet: A deformable network for retinal vessel segmentation," *Knowledge-Based Systems.* **178**, 149-162 (2019).

[5] Gu. Z et al., "CE-Net: Context Encoder Network for 2D Medical Image Segmentation," *IEEE transactions on medical imaging*. 1-1 (2019).

[6] Agrawal. V at al., "Enhanced Optic Disk and Cup Segmentation with Glaucoma Screening from Fundus Images using Position encoded CNNs, " *arXiv preprint arXiv:1809.05216*.

[7] Fu. H et al., "Joint Optic Disc and Cup Segmentation Based on Multi-label Deep Network and Polar Transformation," *IEEE transactions on medical imaging*, **37**(7), 1597-1605 (2018).
14